%% file: slips.tex
\documentclass[sigconf,nonacm]{acmart}

\usepackage{xurl}
\usepackage{booktabs}
\usepackage{tabularx}
\usepackage{array}
\usepackage{makecell}

\newcolumntype{L}{>{\raggedright\arraybackslash}X}
\newcolumntype{C}{>{\centering\arraybackslash}X}

\AtBeginDocument{%
  \providecommand\BibTeX{{%
    \normalfont B\kern-0.5em{\scshape i\kern-0.25em b}%
    \kern-0.8em\TeX}}}

\renewcommand\footnotetextcopyrightpermission[1]{}
\begin{document}

\title{Slips: Behavioral Evidence Aggregation for Network Security}


\author{Sebastian Garcia}
\email{sebastian.garcia@agents.fel.cvut.cz}
\affiliation{%
  \institution{Faculty of Electrical Engineering, Czech Technical University in Prague}
  \country{Czechia}
}

\author{Veronica Valeros}
\email{valerver@fel.cvut.cz}
\affiliation{%
  \institution{Faculty of Electrical Engineering, Czech Technical University in Prague}
  \country{Czechia}
}

\author{Alya Gomaa}
\email{alyaggomaa@gmail.com}
\affiliation{%
  \institution{Faculty of Electrical Engineering, Czech Technical University in Prague}
  \country{Czechia}
}

\author{Ond\v{r}ej Luk\'a\v{s}}
\email{ondrej.lukas@aic.fel.cvut.cz}
\affiliation{%
  \institution{Faculty of Electrical Engineering, Czech Technical University in Prague}
  \country{Czechia}
}

\author{Martin \v{R}epa}
\email{repa.martin@protonmail.ch}
\affiliation{%
  \institution{Recon Wave}
  \country{Czechia}
}

\author{Luk\'a\v{s} Forst}
\email{lukas.forst@gmail.com}
\affiliation{%
  \institution{Recon Wave}
  \country{Czechia}
}

\author{David Otta}
\email{do.dipl.strat@gmail.com}
\affiliation{%
  \institution{Faculty of Electrical Engineering, Czech Technical University in Prague}
  \country{Czechia}
}

\author{Franti\v{s}ek St\v{r}as\'ak}
\email{frenky.strasak@gmail.com}
\affiliation{%
  \institution{Faculty of Electrical Engineering, Czech Technical University in Prague}
  \country{Czechia}
}

\author{Jan Svoboda}
\email{svobo114@fel.cvut.cz}
\affiliation{%
  \institution{Faculty of Electrical Engineering, Czech Technical University in Prague}
  \country{Czechia}
}

\author{Dita Hollmannov\'a}
\email{holl.dita@gmail.com}
\affiliation{%
  \institution{Independent Researcher}
  \country{Czechia}
}

\renewcommand{\shortauthors}{Garc\'ia et al.}

\begin{abstract}
Network intrusion detection systems often analyze individual packets or flows, although malicious behavior may develop across many connections and over time. This may limit their ability to combine isolated detections into a coherent assessment of host behavior. Packet-level features may also be too low-level for complex AI-based detection, requiring additional processing to improve accuracy while maintaining a low false-positive rate.

We present Slips, a network intrusion detection system that builds host-centered behavioral profiles and organizes activity into time windows. It uses a modular architecture in which independent modules report evidence rather than generating final alerts directly. Slips then accumulates this evidence into host-level decisions. We evaluate Slips against Suricata on an expert-labeled PCAP dataset. At the profile-time-window level, Slips achieved 83\% higher recall and a 70\% higher F1 score than Suricata, while neither system produced false positives. These results indicate that time-window-based evidence accumulation can produce context-aware decisions that better align with expert judgment.

\end{abstract}

\ccsdesc[500]{Security and privacy~Intrusion detection systems}
\ccsdesc[300]{Security and privacy~Network security}
\ccsdesc[300]{Computing methodologies~Supervised learning}
\ccsdesc[100]{Computing methodologies~Neural networks}

\keywords{intrusion detection, behavioral analysis, machine learning, evidence ensemble, recurrent neural networks}

\maketitle

\section{Introduction}
Network intrusion detection systems (NIDSs) often analyze individual packets, flows, or protocol events to decide whether activity is malicious. However, malicious behavior may develop across many connections and over time. This leaves NIDSs with limited context for deciding when individual suspicious events indicate that a host is behaving maliciously~\cite{peng_ning, gu_bothunter_2007}.

For example, several failed connection attempts to different ports may be insufficient to determine whether they constitute a port scan. Similarly, command-and-control traffic may be characterized by the periodicity of a sequence of flows rather than by the content of just a single packet~\cite{gu_bothunter_2007}. An accurate detection requires a balance between low-level flow detections and high-level behavioral analysis.

Intrusion detection systems have been studied and deployed for decades, with mature taxonomies, architectures, and operational tools spanning host- and network-based detection, signatures, anomaly detection, and stateful protocol analysis~\cite{axelsson_intrusion_2000,roesch_snort_1999}. Snort and Suricata perform stateful packet and stream inspection and can apply thresholds within individual rules~\cite{roesch_snort_1999,suricata_thresholding}. Zeek converts traffic into protocol events that scripts can relate across connections~\cite{paxson_bro_1999}, while systems such as Kitsune maintain online traffic statistics for an anomaly model~\cite{mirsky_kitsune_2018}.

Graph-based and alert-correlation systems aggregate activity or detector outputs into higher-level reports~\cite{cheung_grids,valeur_comprehensive_2004}. Existing architectures therefore provide some type of state, aggregation, or correlation, but do not combine all three elements presented by our proposal: a shared behavioral profile for each host using time windows, a common representation for evidence produced by diverse detectors, a modularized architecture with many modules and a host-level decision that preserves links to the traffic supporting it. We compare these architectural differences in Section~\ref{sec:previous-work-architectures}. 

We describe Slips~\cite{anonymous_slips}, first created in 2012\cite{anonymous_disertation}, a modern modular network security system that uses individual detections to analyze host behavior over time.
Although Slips consumes network traffic and performs intrusion detection and prevention, its architecture is closer to a Security Orchestration, Automation, and Response (SOAR) system than to an IDS~\cite{nist_soar}. It collects and normalizes network observations, maintains shared state, correlates findings from diverse modules, exports to different systems, makes a higher-level decision, and can act back on the network to stop an attack.

Slips models each IP address as a behavioral profile divided into configurable time windows. AI and non-AI modules contribute evidence about the profile’s flows, protocol events, and derived behavior; Slips accumulates this evidence before raising alerts and preserves the contributing detections and supporting flows for accountability, as described in Section~\ref{sec:input}.

This paper focuses on Slips’ core architecture, how organizing detections by profile and time window improves detection, and how low-level detections are combined into alerts when sufficient evidence is available.


To evaluate Slips, we conducted two sets of experiments, described in Section~\ref{sec:evaluation}. First, we ran Slips and Suricata on three PCAP files containing port scans of different sizes and durations (100, 1000, and 65,536 ports). The goal of this experiment is to evaluate whether each IDS can identify basic scan-related behavior, how Slips accumulates evidence before raising an alert, and what level of activity is required before each system reports it. A port scan is one of the most basic activities that an IDS should detect, yet to this day, it remains surprisingly difficult to achieve correctly.

Second, we ran Slips and Suricata on three different PCAP files, two malicious (Bladabindi RAT and TrickBot) and one benign (social media traffic on Windows 7), and evaluated the performance metrics of each system using the expert-labeled ground-truth log files for each PCAP. These experiments use two comparison modes: flow-by-flow, which supports the way Suricata works, and profile-window, which supports the way Slips works. The goal of this experiment is to compare signature-based Suricata with behavioral- and time-window-based Slips in terms of their detection performance metrics.


Results of the port scan experiments show that Slips generates scan-related evidence for all three scan captures. It confirms that Slips separates the generation of intermediate evidence from the final alerting: suspicious behavior can be recognized and stored without being escalated immediately. Suricata did not recognize any of the three port scans and raised no alerts.

Results from the malware and benign experiments show that Slips achieved stronger detection performance than Suricata across both evaluation modes, while maintaining a very low false-positive rate. In the flow-by-flow mode comparison, Slips achieved an F1 score of 0.1603 compared with 0.0033 for Suricata, with both systems producing the same false-positive rate of 0.000025. In the profile-window view, Slips achieved an F1 score of 0.3268, compared with 0.1925 for Suricata, and both systems produced zero false positives. These results suggest that accumulating evidence across behavioral profiles and time windows can improve host-level decision-making while avoiding unnecessary alerts from isolated detections.


This paper makes seven contributions.
\begin{enumerate}
\item Modular architecture for network security analysis in which detectors, ML models, exports, and responses are connected through shared behavioral states.

\item A description of host profiles based on time windows as the main analysis entity.

\item An adaptive evidence ensemble method in which diverse modules, including an online ML flow classifier and a multi-flow ML recurrent model, produce accountable evidence as behavior evolves, and of how Slips accumulates that evidence into host-level alerts.

\item A mechanism that allows end users to adapt and extend the training of ML models using their own traffic.

\item Organization-level whitelisting that allows users to exclude alerts and block requests to or from domains, IPs, and certificates across multiple protocols.

\item Comparison between Slips and Suricata using PCAP files from the IDSEVAL dataset.

\item New dataset to evaluate IDS systems, including a new labeling framework and a new performance comparison tool.
\end{enumerate}

\section{Slips System Overview}
\label{sec:overview}

\begin{figure}[t]
  \centering
  \includegraphics[width=\linewidth]{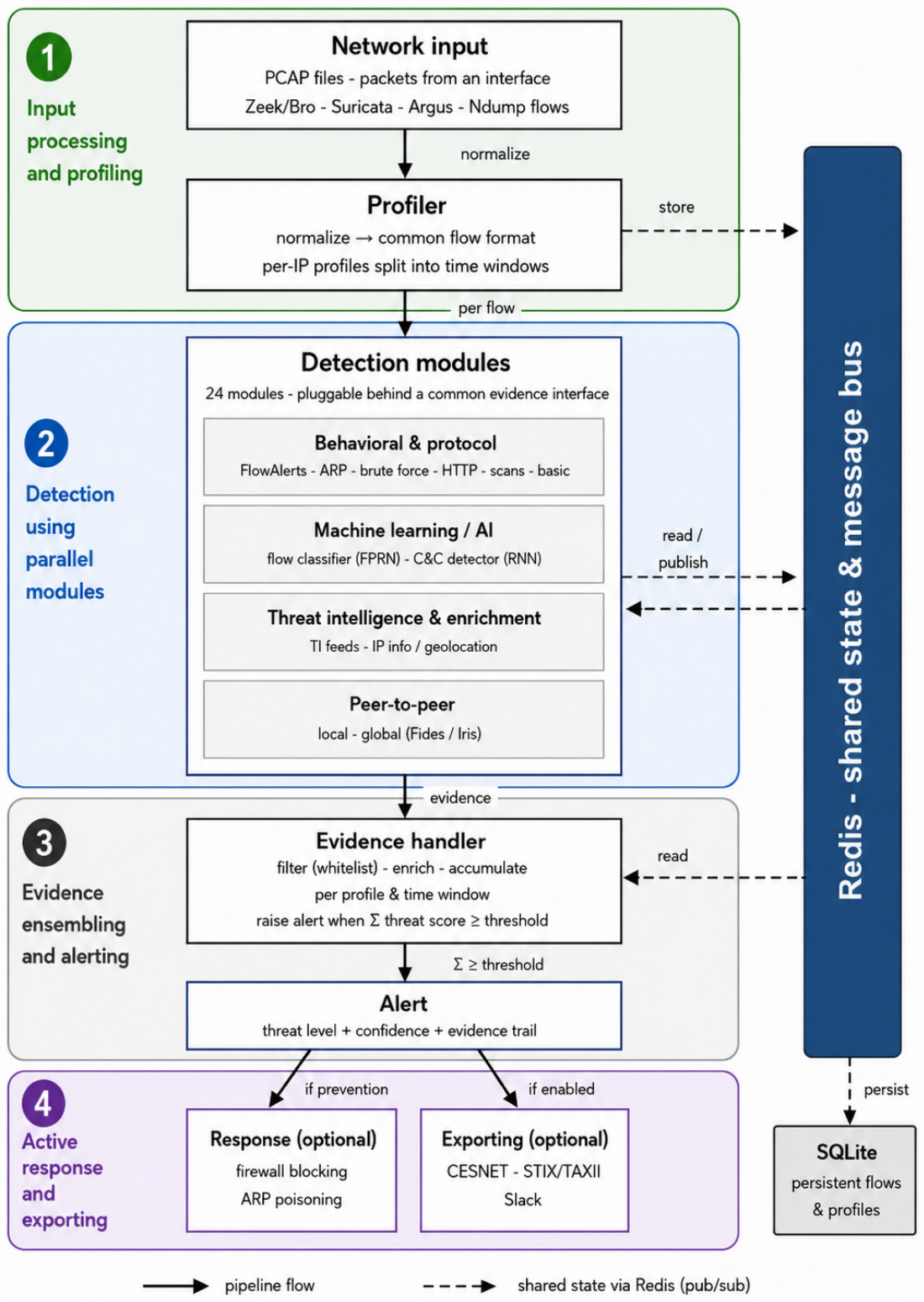}
  \caption{Core Slips architecture. (1) Network inputs are normalized into a common flow format and aggregated into host profiles and time windows. (2) Detection modules analyze the input data to emit evidence. (3) Evidence is analyzed as an ensemble, and alerts are decided. (4) Responses are executed, and data is exported. Redis provides a pub/sub architecture for communication between parts and for runtime data, while SQLite provides persistent storage.}
  \label{fig:architecture}
\end{figure}

The core Slips architecture, shown in Figure~\ref{fig:architecture}, consists of four major parts that are continually running in parallel: (1) input processing and profiling, (2) detection modules, (3) evidence ensembling and alerting, and (4) active response and exporting. 

In the input processing and profiling part, described in Section~\ref{sec:input}, network traffic enters through the input process in various formats, such as PCAP and Zeek logs, among others. The profiler then converts the input traffic into a common internal flow format, creates the host profiles and time windows, stores the resulting records in the database, and dispatches the flows to the modules using a pub/sub architecture.

In the detection modules part, described in Section~\ref{sec:detection-modules}, each module analyzes the input data to generate evidence. Evidence is a detection related to one or more flows.

In the evidence ensembling part, described in Section~\ref{sec:ensemble}, the evidence handler receives evidence from each module, filters it using the whitelist, and enriches it. When sufficient evidence accumulates in the ensemble, the evidence handler raises an alert and may request an active response.

In the active response and exporting part, described in Section~\ref{sec:active-response}, the blocking module blocks the detected attacker through the firewall, isolates the local-network attacker using ARP poisoning, and exports alerts.

\subsection{Input Normalization and Profiling}
\label{sec:input}
Slips can ingest packets from network interfaces and many network traffic formats, including PCAP and PCAP-NG~\cite{ietf_pcap,ietf_pcapng}, Zeek logs~\cite{zeek_logs}, Suricata EVE logs~\cite{suricata_eve}, Argus BinetFlow binary logs~\cite{openargus}, and nfdump flows~\cite{nfdump}.

Packet captures and live traffic are processed by a Zeek process controlled by Slips and then ingested, while the rest of the input formats are specifically parsed. All input are converted to a unique common flow format for all modules.

In the case of using a network interface or PCAP file, Slips uses an extended Zeek with eleven custom scripts that add fields and logs, including: JA3 and JA3S TLS fingerprints~\cite{salesforce_ja3_ja3s}, ARP events, ICMP scan events, DNS-over-HTTPS signals, SHA256 file hashes, network gateway information, IRC protocol, and TELNET, rlogin, and rsh login records.

After the input data is processed, the profiler code creates a \textit{behavioral profile} structure for each observed source IP address and partitions its activity into time windows (by default, 1 hour). This means that each profile-time-window (from now on, just profile) groups all the activity of that source IP. This lets Slips analyze behavior as a continuous process rather than just treating a host as \textit{infected or not}. New flows are assigned to the profile by timestamp, including late or out-of-order arrivals. A profile, therefore, represents host behavior over time rather than a single connection.

Slips input processing can operate in two analysis-direction modes: \texttt{out} and \texttt{both}. The \texttt{out} mode answers the question "Is my computer infected and communicating maliciously?" by storing only \textit{outgoing} connections from each profile. The \texttt{both} mode answers the question "Is my computer also being attacked?", by storing both outgoing and incoming connections on each profile. 

\subsubsection{Threat Intelligence}
\label{sec:threat-intelligence}
Slips includes a feed update manager that periodically downloads, normalizes, and refreshes several types of threat intelligence (TI) feeds. These include malicious JA3 and JA3S fingerprints, SSL certificate fingerprints, Tor exit nodes, the Tranco top benign domains~\cite{tranco}, and 42 public TI feeds\footnote{\url{https://anonymous.4open.science/r/Slips-731B/config/TI_feeds.csv}}. The data are used as blacklists or whitelists, and also enrich IP addresses referenced in evidence.

The IP info module, described later, also adds metadata to each IP address, including ASN, geolocation city (using GeoLite DBs), WHOIS data, domain creation date, domain registrant information, and MAC address vendor. It also uses a heuristic to infer the local gateway's IP and MAC addresses. Lookups run asynchronously alongside modules, and the retrieved metadata are attached to the corresponding evidence before reporting. All feeds and auxiliary databases are locally cached and refreshed periodically by configuration.

\subsubsection{Whitelists}
\label{sec:whitelists}

Slips supports user-defined and built-in whitelists to reduce context-dependent false positives and suppress known benign activity. For example, traffic from an internal monitoring server may resemble scanning but be expected in a particular network.

\paragraph{User-Defined Whitelist}

Each whitelist entry specifies an \texttt{indicator type}, value, \texttt{direction}, and \texttt{ignore mode}. Supported indicator types are IP addresses, domains, MAC addresses, and organizations. The direction determines whether the entry applies to the flow source, destination, or both, while the ignore mode allows ignoring the entry for flows, alerts, or both.

In flow-ignoring mode, Slips drops matching flows before enrichment, detection, storage, and display in the web interface. This reduces processing overhead and removes known benign traffic from the analyst's view. However, information from these flows is unavailable to later detections. For example, whitelisting a DNS server in this mode prevents Slips from learning its role and hides its traffic from the interface.

In alert-ignoring mode, matching flows remain available for enrichment, detection, and display, but their evidence and alerts are suppressed. Any generated evidence is discarded before alert accumulation and, therefore, cannot trigger an alert or blocking request. For each evidence item, Slips checks both the attacker and victim against whitelisted IP addresses, DNS resolution answers, queried domains, CNAMEs, TLS SNI values, URLs, and MAC addresses.


Organization-aware whitelisting is a novel feature of Slips that extends conventional IP and domain-based whitelists. Slips allows users to whitelist entire organizations by putting their names, such as Google, Microsoft, Apple, Facebook, and Twitter. Slips expands these organizations by querying their known domains, IP ranges, and ASNs.

A user can whitelist traffic associated with an organization and configure whether matching flows, alerts, or both are suppressed. This helps remove known benign organizational traffic from the analyst’s view and keeps attention focused on more relevant activity.

\paragraph{Built-In Whitelist}

In addition to user-defined entries, Slips optionally can use the top 10,000 domains from the Tranco list, a daily ranking of the one million most popular websites~\cite{tranco}. The feed update manager downloads and caches the list daily. This list helps prevent false positives involving well-known destinations. For example, a domain flagged by a threat-intelligence source may be ignored when it belongs to a known whitelisted organization such as Google. 

Tranco domains are whitelisted only at the alert level: related flows remain available for processing and inspection, but their evidence is suppressed and does not contribute to alerts.

\subsection{Detection Modules}
\label{sec:detection-modules}

Slips currently has 24 independent modules, including behavioral and ML-based detectors. Each module runs as a separate process and has access to both Redis and SQLite through a shared database interface. Each module registers in the Redis channels for the type of flows that it needs, for example, \textit{DNS Request}.

Each detection module focuses on a specific behavioral family and emits evidence when it finds suspicious activity. Evidence is a detection that may include one or more flows and is sent to the evidence handler rather triggering a response. 

Most modules use heuristic behavioral and adaptive rules rather than stateless ones. They maintain a state for each profile and time window, and update their detections as new events arrive. As a result, their decisions depend on the number and timing of related events, rather than on a fixed property of a single packet or flow.



Detection modules can be enabled, disabled, or added from the configuration file, and some modules allow individual detections to be disabled without disabling the entire module. Based on their purpose, detection modules are grouped into behavioral, peer-to-peer, machine-learning, defense, and export modules.

Each evidence generated by a module must have a threat level (Info, Low, Medium, High, and Critical) and confidence (from 0 to 1). The Info threat level indicates they are worth reporting, but do not pose a direct threat so they don't count towards the alert score. Critical evidence contributes significantly to the score required to generate an alert.

\subsubsection{Behavioral Modules}

Behavioral modules analyze specific network protocols to detect suspicious activity. They cover both single-event anomalies (e.g., malformed or unexpected protocol use) and adaptive, event-count-based behaviors (e.g., scans and brute-force attacks). 

\paragraph{Flow Alerts}
The Flow Alerts module is a dispatcher for a collection of protocol-specific analyzers. It consumes normalized flow records and selected Zeek logs for DNS, TLS/SSL, SSH, SMTP, tunnels, software, Zeek notices, and remote login activity (e.g., Telnet, rlogin, and rsh), as well as information from downloaded-file logs.

This module contains the majority of Slips detections. It flags long connections, unknown destination ports, repeated rejected connections, repeated Telnet attempts, data uploads, TOR exit-node connections, and connections made without a prior DNS resolution.

Several Flow Alert detections adapt their behavior based on event counts. For example, repeated rejected connections, repeated Telnet attempts, SMTP brute-force, bursts of NXDOMAIN replies, and repeated empty HTTP connections are evaluated based on the events accumulated for the profile within the time window. These detections, therefore, depend on the number and timing of related events, not on any single flow.

For DNS flows, the Flow Alerts module detects young domains, DNS answers with high-entropy TXT records, private IPs in DNS answers, many NXDOMAIN replies that may indicate DGA behavior, ARPA scans, and DNS resolutions that are not followed by a connection.

For TLS/SSL flows, it detects self-signed certificates, malicious JA3 or JA3S fingerprints, certificate common-name mismatches, suspicious organization names in certificates, DNS-over-HTTPS, large Pastebin downloads, and non-SSL traffic on port 443.

For SSH flows, it detects successful SSH logins. It uses Zeek's \texttt{auth\_success} field when available; otherwise, it estimates success from the total bytes exchanged.

For SMTP flows, it detects invalid SMTP logins and SMTP brute-force attempts based on Zeek fields.

It also detects GRE tunnels and GRE scans, multiple uses of SSH client or server versions, malicious TLS/SSL certificates from downloaded files, and login events from Zeek's \texttt{login.log}.

\paragraph{ARP}
The ARP module detects attacks and abnormal ARP behavior in the local network. It raises evidence for:
\begin{itemize}
  \item ARP scans, when one host sends ARP requests to at least five different IP addresses within 30 seconds,
  \item ARP packets sent outside the configured local network, and
  \item Unsolicited ARP replies.
\end{itemize}

The ARP scan detector is adaptive over a short time interval: one ARP request is normal, but a burst of requests to several different addresses within 30 seconds is treated as discovery behavior.

It also checks for MITM ARP attacks. If the same MAC address was previously linked to one IP and later appears to claim another IP, Slips raises evidence of a possible ARP cache poisoning attack.

\paragraph{Brute-force Detector}
The brute-force detector focuses on detecting SSH brute forcing. It is adaptive: for each source profile, time window, destination IP, and destination port, it aggregates failed SSH sessions and authentication attempts into a campaign. When the number of failed logins reaches the configured threshold, Slips raises password-guessing evidence. After the threshold is reached, it does not emit the same evidence for every new attempt; instead, it reports again only at sparse, bucketed points as the campaign grows. Confidence increases with the number of attempts and reaches full confidence after a larger number.

The module also uses SSH client banners. Tools and libraries such as Hydra, Medusa, and Ncrack are purpose-built for automated authentication testing, while Paramiko and \texttt{libssh} enable programmatic SSH sessions~\cite{hydra_github,medusa_github,ncrack_official,paramiko_project,libssh_project}. Their presence, therefore, increases the confidence used in the evidence.

\paragraph{HTTP Analyzer}
The HTTP analyzer checks HTTP-specific behavior and raises evidence for:
\begin{itemize}
  \item Suspicious user agents, abrupt user-agent changes per profile, and mismatches between the user-agent OS and the MAC vendor,
  \item Executable downloads inferred from HTTP response MIME types,
  \item Repeated empty HTTP connections to common sites (a pattern that can be consistent with malware checking internet connectivity)~\cite{mitre_internet_connection_discovery},
  \item Large Pastebin downloads above a configured size threshold,
  \item Unknown HTTP methods reported by Zeek's \texttt{weird.log},
  \item Established TCP traffic on port 80 that Zeek did not classify as HTTP.
\end{itemize}

\paragraph{Leak Detector}
The leak detector searches packet captures for sensitive data patterns and runs only when Slips analyzes a PCAP. It applies YARA rules shipped with the module (including GPS-location leak patterns)~\cite{yara_official}.
When a rule matches, Slips uses \texttt{tshark}~\cite{tshark} to extract the source and destination IPs, protocol, ports, and timestamp, and reports them as evidence.

\paragraph{Network Discovery}
The network discovery is an adaptive module that detects scanning and probing behavior. It raises evidence for:
\begin{itemize}
  \item vertical port scans (one source contacting many ports on the same destination IP using non-established TCP or UDP flows),
  \item horizontal port scans (one source contacting the same port on many destination IPs),
  \item ICMP sweeps reported by Zeek notices (including timestamp scans, address scans, and address-mask scans), and
  \item DHCP scans, when a client requests four or more different IP addresses in the same time window.
\end{itemize}

The main problem of detecting port scans is that tools can not generate a new alert for each new port scanned, but also they need to show the difference between scanning 5 ports and 65,536. Therefore, Slips uses a logarithmic scale to detect port scans: it emits evidence when the scan grows enough to cross a new threshold of behavior, rather than setting an evidence for every small increase. The confidence is determined by the volume of supporting traffic, so the evidence score increases as more packets and destinations are observed.

\subsubsection{Peer-to-Peer}

Slips is the first IDS to implement a peer-to-peer network of detectors both in the local network and globally on the Internet by using the libP2P library. If two computers with Slips are on the same local network, they will find each other and start sharing data. The goal is to share detection information so as to better protect each other. 

To protect privacy, Slips peers only share the attacker’s IP address or domain, threat score, and confidence value, rather than full traffic or contextual data.

\paragraph{Local P2P and Local Trust}
The local P2P module enables Slips instances on the same network to ask each other for an IP address and to share detections. The module queries local peers, waits briefly for responses, aggregates peer scores using a trust model, and stores the resulting network opinion.

Slips raises evidence when the aggregated peer opinion indicates that the IP is suspicious. In addition to queries, the local P2P module receives blame reports when another peer blocks an attacker. The local trust model is inspired by the trust model of the Sality P2P botnet~\cite{falliere_sality_2011}: trust depends on good communication and on how long peers have known each other, with peers unable to influence the trust other peers have in them because each peer computes trust locally rather than asking others for it. This is crucial to avoid adversarial peers manipulating the network.

\paragraph{Global P2P and Global Trust}
Slips also has a global P2P module that implements three DNS-based supernodes on the Internet, allowing every public Slips instance to automatically share and receive threat intelligence. It also allows organizations to define their own private feed of peers using cryptographic signatures.

Peers respond with an opinion represented as a score and confidence. To avoid blindly trusting adversarial responses, Slips applies a trust model, aggregates peer reports, and then caches the final network opinion. As with local P2P, shared information is minimal and does not include full evidence contents or private flow details.

\subsubsection{Machine Learning and AI}

The ML modules in Slips are a crucial and hard part to maintain, since the models are continually retrained using our own datasets and pipelines. Before publication, we compare many models in large datasets to decide which model to ship. ML modules operate at two levels: an online linear classifier to detect individual flows, and a bidirectional gated recurrent unit (GRU) to detect command-and-control behavior. Both models are free, open, and distributed with Slips.

\paragraph{Flow-Level ML Detection}

The flow-level ML detection module applies a pre-trained \texttt{SGDClassifier} and \texttt{StandardScaler} to each eligible normalized flow. The shipped model is trained on our own curated collection of security datasets, including the public Security Datasets for Testing collection~\cite{security_datasets_testing}. The module removes unused fields, excludes protocols without the required port-based representation, encodes transport protocol and connection state, and uses port, duration, packet-count, and byte-count features.

During normal operation (test mode), flows are analyzed individually, and only high-confidence detections are sent as evidence. Since the difference between the training datasets and the user's traffic is usually high, the default threat level is low, and confidence is also low at 0.1.

More importantly, the module allows end users to extend the ML model locally using their own traffic. This module implements a type of transfer learning for this purpose. The user can set the learning mode in the configuration to \texttt{train} and the default label to \texttt{normal} or \texttt{malicious}, and then run Slips normally. The user-defined label is then applied to the flows, and the module continues training the shipped model with the new data. The updated classifier and scaler are stored on disk. By switching the mode back to \texttt{test} the user can now use an extended and probably better model.

This mechanism transfers a model trained on distributed data to the operator's local environment without requiring a separate ML pipeline or model replacement. Technically, it is an incremental supervised adaptation of a linear model, rather than neural transfer learning through frozen and fine-tuned layers. This is the first time, as far as we know, of a ML model that allows the end user to adapt it locally.

\paragraph{Recurrent Command-and-Control Detection}

The ML command-and-control detection module analyzes behavior across network flows to detect command-and-control channels. For each profile, the module receives flows that are aggregated by the source IP address, destination IP address, destination port, and protocol. All these flows, aggregated, form a tuple that represents the behavior of the source IP towards a specific service. Slips converts each flow in this tuple into a letter (e.g., 'a'), which discretizes the flow duration, size and periodicity (relative to flows). An additional letter is used to encode the elapsed time since the previous flow (e.g., '*'). The resulting string of letters is analyzed by the module's ML model in search of well-known behavioral patterns in command and control channels.

The current model is a bidirectional GRU distributed as a pre-trained Keras model~\cite{schuster_brnn_1997,cho_rnn_encoder_decoder_2014}. The shipped model is trained on our own curated collection of security datasets, including the public Security Datasets for Testing~\cite{security_datasets_testing}. 

A command-and-control detection requires a model score greater than \(0.99\). Confidence is not the raw model score: it grows with observed sequence length as \(\min(n/100,1)\), where \(n\) is the number of encoded characters, so short matches contribute less than behavior supported by a longer history.

When the threshold is crossed, the module creates evidence with a high threat level (\(0.8\)) and reports the flow that triggered the evaluation with the evidence. Each item then enters the ensemble for its own profile and time window. The GRU therefore detects a multi-flow temporal pattern.

\subsubsection{Exporting Modules}

Slips provides exporting modules that exchange IoCs, evidence, and alerts with external systems. Internally, Slips records each evidence item and alert as JSON in the IDMEFv2~\cite{idmefv2} format in \texttt{alerts.json}. Evidence items are represented as IDMEFv2 \texttt{Event} objects, while alerts are represented as IDMEFv2 \texttt{Incident} objects. In addition to this internal representation, Slips can export alerts to external platforms in several formats, including STIX over TAXII~\cite{stix,taxii}, CESNET Warden~\cite{cesnet_warden}, and Slack notifications.

\paragraph{Exporting Alerts Module}
The exporting alerts module sends Slips evidence and alerts to external systems.

It supports exporting alerts to Slack and sending STIX-formatted evidence to TAXII servers. Slack alerts are posted to a configured Slack channel through a bot token, while STIX export converts evidence into STIX indicators before transmitting them to the configured TAXII server.

\paragraph{CESNET Module}

The CESNET module shares alerts with the Warden servers operated by the CESNET organization~\cite{cesnet_warden}. When CESNET export is enabled, the module converts evidence into the IDEA0 format~\cite{cesnet_idea} and sends it to a Warden server. Informational evidence and evidence about local IP addresses are not exported. The module can also receive alerts from Warden servers, convert them into threat intelligence data, and use them as a blacklist for future detections.

\subsection{Evidence Ensemble and Alerting}
\label{sec:ensemble}

Slips combines evidence from different detection methods. It does not average model probabilities or require its detectors to share features. Instead, every rule, ML model, behavioral detector, and threat-intelligence module converts its result into an \emph{evidence} object and sends it to the evidence handler process for processing. Each evidence item includes the detection type, attacker, and optional victim, profile, time window, threat level, confidence, timestamp, detection method (e.g., AI or behavioral), relevant protocol and ports, and identifiers of the supporting flows. Related evidence can also be linked.

The evidence handler validates each new evidence item before it is accumulated. It applies the configured whitelist and ignores informational evidence when computing the score. Evidence can still be logged, exported, or shared independently of whether it raises the accumulated score.

For decision-making, Slips maps \emph{low}, \emph{medium}, \emph{high}, and \emph{critical} threat to \(0.2\), \(0.5\), \(0.8\), and \(1.0\), respectively. Evidence \(i\) contributes \(s_i=t_i c_i\), where \(t_i\) is this threat value and \(c_i\in[0,1]\) is detector confidence. The ensemble score for profile \(p\) and time window \(w\) is \(S_{p,w}=\sum_i s_i\) over eligible evidence assigned to that profile and window. Informational evidence contributes zero, whitelisted evidence is removed, and evidence already used in a previous alert is not counted again.

When \(S_{p,w}\) reaches the configured alert threshold, the handler creates an alert containing the profile, time window boundaries, accumulated score, and identifiers of all contributing evidence. Thus, a high-confidence GRU detection can contribute strongly, a low-confidence or flow-level ML result contributes less, and several independent ML and non-ML findings can jointly cross the threshold. No individual model is allowed to make the final alerting decision on its own.

This separation creates three explicit semantic levels. Flows and protocol events are observations; evidence is a module's accountable interpretation of one or more observations; and an alert is the system's combined decision about a host over a period. The links among these levels allow an analyst or a downstream system to reconstruct which modules contributed, how strongly, and which traffic supports the alert.

\subsection{Active Response and Exporting}
\label{sec:active-response}

Slips active response includes logging alerts and evidence, optionally exporting them to external systems, and optionally triggering a defense module.

Alerts and evidence are persisted and exposed through the command-line and web frontends, with the native \texttt{alerts.json} log written in IDMEFv2~\cite{idmefv2} as described above. 

The exporting alerts module can send notifications to Slack, send evidence to a TAXII~\cite{taxii} server, or exchange IDEA events through CESNET Warden~\cite{cesnet_warden}.

When prevention is enabled, Slips uses its defense modules to block traffic to and from an attacker through the host's iptables firewall~\cite{netfilter_iptables}.
On a local network, Slips can instead use ARP poisoning as a defense mechanism to block all traffic to and from the detected attacker. This happens in two steps: (1) isolating the attacker from the gateway, which makes the attacker unable to access the internet, and (2) disrupting other local hosts' connections to and from the attacker, which makes the attacker unreachable. 

Both defense mechanisms include automatic unblocking. After being blocked, a profile enters a probation period defined in time windows. New alerts extend this period, while a profile that generates no additional evidence is automatically unblocked when the probation period ends.

\section{Evaluation Methodology}
\label{sec:evaluation}
The goal of the evaluation is to show that the current Slips architecture can produce good detections similar to SOTA IDS. The evaluation contains two sets of packet captures with different purposes. The first set contains three controlled vertical port-scan captures and evaluates whether the architecture represents the behavior distributed across flows as accountable evidence. The second set contains malware and benign captures and compares the detection decisions of Slips and Suricata. Table~\ref{tab:evaluation-datasets} summarizes the six captures used in the two experiment sets.

Suricata is used here because it is a widely deployed, mature, feature-rich IDS that represents a strong production baseline rather than a one-off research method.

These datasets are intentionally illustrative. They are not meant to represent the full detection coverage of either Slips or Suricata. Both systems can detect many behaviors that are not present in these captures, so the evaluation should be read as an architectural demonstration with a small controlled comparison, not as a benchmark of general IDS capability.

\begin{table*}[t]
\centering
\caption{Evaluation datasets used to assess evidence generation for vertical port scans and to compare Slips and Suricata on malware and benign user traffic.}
\label{tab:evaluation-datasets}
\scriptsize
\begin{tabular}{cllll}
\toprule
Experiment set ID & PCAP ID & Capture & Traffic content & Evaluation purpose \\
\midrule
1 & Portscans-1 & \texttt{idseval-malicious-portscan-1.pcap} & Vertical port scan (100 ports) & Evidence generation \\
1 & Portscans-2 & \texttt{idseval-malicious-portscan-2.pcap} & Vertical port scan (1,000 ports) & Evidence generation \\
1 & Portscans-3 & \texttt{idseval-malicious-portscan-3.pcap} & Vertical port scan (65,536 ports) & Evidence generation \\
2 & Malware-1 & \texttt{idseval-malicious-malware-1.pcap} & Bladabindi RAT & Performance comparison \\
2 & Malware-2 & \texttt{idseval-malicious-malware-2.pcap} & TrickBot malware & Performance comparison \\
2 & Benign-1 & \texttt{idseval-benign-user-traffic-1.pcap} & Social media browsing & Performance comparison \\
\bottomrule
\end{tabular}
\end{table*}

\subsection{Ground Truth and Expert Labeling}

Domain experts establish the ground truth before running either detection system. They inspect the capture context, identify the malicious hosts, attack intervals, and relevant traffic, and encode those judgments as labeling conditions.

We then apply the open-source NetFlowLabeler tool~\cite{anonymous_netflowlabeler} which applies the expert-authored conditions to individual flows and produces generic and detailed labels. The tool makes the labeling procedure repeatable; it does not replace the expert judgment used to define the conditions. 

For binary evaluation, flows labeled as malicious form the positive class. Flows labeled as benign or background form the negative class. The artifacts accompanying this paper~\cite{anonymous_dataset_2026} include the PCAP files (IDSEVAL dataset), labeling configurations, and the cryptographic hashes.

\subsection{Experiment Set 1: Port-Scans}

The first experiment uses three PCAPs containing known port scans (done using the nmap tool) and evaluates both Slips and Suricata to understand whether each detects them and how each tool reports them.
For Slips, each capture was processed with Slips v1.1.21, a one-hour time window, an evidence threshold of 15, and the default module configuration. We then report, for each capture, whether Slips generates any evidence of vertical or horizontal scans. We record whether evidence was generated, the accumulated evidence score, and whether the evidence triggered an alert.

For Suricata, each capture was processed by Suricata v8.0.4 (RELEASE) with the Emerging Threats Open ruleset 8.0.4\footnote{Ruleset download: \url{rules.emergingthreats.net/open/suricata-8.0.4/emerging.rules.tar.gz}.} and the default configuration. We then report, for each capture, whether Suricata produced a corresponding scan alert.

In this experiment, the comparison is descriptive because the systems expose different intermediate objects: Slips generates evidence and alerts, whereas Suricata generates rule alerts.

Portscans-1 to Portscans-3 contain only port-scanning traffic, so they cannot be used to measure overall accuracy but only recall. In particular, we cannot tell how often the systems would raise scan alerts on benign traffic because this dataset does not include a defined set of “non-scan” examples. Also, the labelling was done flow by flow, but the concept of \textit{what is a port scan} is left to the IDSs.

Instead, this experiment checks how Slips behaves in comparison with Suricata by asking whether it can (1) recognize scanning patterns that span many connections, and (2) keep a clear trail showing which connections led to that conclusion.

\subsection{Experiment Set 2: Malware and Benign Traffic}
The second experiment evaluates detection performance using two complementary traffic groups: malicious malware captures and benign user traffic. The malicious group consists of Malware-1 and Malware-2, while the benign group consists of Benign-1. Malware-1 contains Bladabindi RAT traffic and Malware-2 contains TrickBot traffic~\cite{microsoft_bladabindi,cisa_trickbot}. Both Slips and Suricata process the same captures. 

Slips first converts each PCAP into Zeek flows and then makes decisions per host profile and time window. 
Suricata processes the original PCAPs directly and reports rule-based alerts.

We compare Slips and Suricata from two perspectives: the flow level and the profile-time-window level. At the flow level, a Slips prediction is considered malicious when the flow identifier is linked to evidence. A Suricata prediction is considered malicious when one of its alerts can be mapped to that flow. The flow-by-flow level was chosen because it is how Suricata works by default. 

For the benign capture, \verb|Benign-1|, any detected flow is counted as a false positive because the PCAP contains benign social media browsing traffic and no malicious activity. All remaining undetected benign flows are counted as true negatives.

In addition to the flow-level comparison, we also evaluate both tools at the profile-time-window level because this corresponds to the native decision unit used by Slips. In the ground truth, a profile-time-window is considered malicious if it contains at least one malicious flow in that profile-time-window. 

Slips directly provides predictions at this level, whereas for Suricata, alerts are post-processed and grouped by IP address and assigned to the same time-window boundaries used by Slips.

This profile-time-window-level comparison evaluates whether each system can identify malicious host behavior as it develops over time, rather than only detecting individual malicious flows in isolation.

\subsection{Metrics and Statistical Reporting}

For Malware-1 and Malware-2 datasets, the primary quantities are true positives and false negatives, from which we compute recall as \(TP/(TP+FN)\). For Benign-1, we report false positives and true negatives and compute the false-positive rate as \(FP/(FP+TN)\). Across the combined malware and benign sets, we additionally report precision, specificity, F1 score, and the complete confusion matrix for both systems.

Metrics are reported separately for the flow-level and profile-time-window evaluation views. Per-capture results are provided in the accompanying artifacts, while the results section reports aggregate metrics computed using micro-aggregation.

\subsection{Systems, Controls, and Reproducibility}

For every run, we record the Slips and Suricata versions, enabled Slips modules, Suricata ruleset, configuration files, time window width, evidence threshold, analysis direction, command lines, and execution environment. 

Slips and Suricata receive identical PCAPs. 

Detection outputs for each system are generated before the comparison tool reads them, and ground-truth labels are not available to either system during detection.

The reproducible pipeline has four stages. First, Zeek converts each PCAP into zeek logs. Second, NetFlowLabeler applies the expert-authored labeling configuration to produce the flow-level ground truth. Third, Slips and Suricata process the capture independently. Fourth, the IDPS comparison tool\cite{anonymous_idps_comparison_tool} maps evidence and alerts with the labeled flows, constructs the profile-time-window view, and computes the selected metrics.

The artifact accompanying this paper~\cite{anonymous_dataset_2026} contains all mentioned PCAPs, checksums, labeling configurations, the expected output of each tool used, and the steps used to conduct the two experiment sets.

\section{Results and Discussion}
\label{sec:results}

\subsection{Experiment Set 1: Port-Scan Evidence Generation Results}

For Portscans-1, Slips generated two scan evidence items and raised no alerts because the accumulated evidence did not reach the configured alert threshold. For Portscans-2, Slips generated one scan evidence item and also raised no alerts. Finally, for Portscans-3, Slips generated five evidence items, but raised no alerts.

For Slips, this is the expected behavior when dealing with portscans: Slips recognizes the port-scan behavior and stores evidence items, but it does not treat one isolated scan as sufficient for a host-level alert. In the current configuration, Slips needs more evidence before deciding to act against that IP address. However, the port scans are detected correctly for the analyst. 

In contrast, Suricata raised zero alerts across Portscans-1--Portscans-3. This is surprising, since a port scan should at least acknowledge the security analysts.

The results are shown in Table~\ref{tab:portscan-results}. The table presents the evidence generation, accumulated score, alert thresholding, and alert generation instead of a confusion matrix because our goal here is not to compare the accuracy of the two systems, but to demonstrate how each tool deals with attacks that happen over time.

The important distinction in this experiment is therefore not only whether an alert is raised, but whether the system exposes an intermediate representation showing that the scan was understood. In Slips' case, the intermediate representation is evidence. In Suricata's case, there is none.

The evidence generated by Slips and their links to supporting flows are available in the \texttt{alerts.json} file included in the Slips output artifacts.

\begin{table*}[t]
\centering
\caption{Port-scan evidence and alert-threshold behavior on the IDSEVAL scan captures. The table reports whether scan-related evidence was generated, the accumulated score for the scanner profile during the experiment time window, the configured alert threshold, and the resulting alert decision.}
\label{tab:portscan-results}
\small

\begin{tabular}{llcccc}
\toprule
ID & IDS & Evidence Generated & Accumulated Score & Alert Threshold Per Hour & Alerts Generated  \\
\midrule
Portscans-1 & Slips     & Yes   & $2.85$    & $15$  & No \\
Portscans-1 & Suricata  & No    & --        & N/A   & No \\
\\
Portscans-2 & Slips     & Yes   & $1.3$     & $15$  & No \\
Portscans-2 & Suricata  & No    & --        & N/A   & No \\
\\
Portscans-3 & Slips     & Yes   & $7.99$    & $15$  & No \\
Portscans-3 & Suricata  & No    & --        & N/A   & No \\
\bottomrule
\end{tabular}
\end{table*}

\subsection{Experiment Set 2: Malware and Benign Traffic Results}



As shown in Table~\ref{tab:malware-results}, at the flow level, Slips produced 3,610 true positives and 37,834 false negatives, resulting in a recall of 8.71e-2. It also produced 2 false positives and 78,740 true negatives, yielding a precision of 9.994e-1, an FPR of 2.5e-5, and an F1 score of 1.603e-1. Suricata produced 68 true positives and 41,376 false negatives, resulting in a substantially lower recall of 1.6e-3 and an F1 score of 3.3e-3. It produced the same 2 false positives and 78,740 true negatives, with a precision of 9.714e-1 and an FPR of 2.5e-5. Thus, Slips achieved substantially higher malicious-flow coverage without increasing the number of false positives.

At the profile-time-window level, Slips produced 33 true positives and 136 false negatives, achieving a recall of 1.953e-1 and an F1 score of 3.268e-1. Suricata produced 18 true positives and 151 false negatives, yielding a recall of 1.065e-1 and an F1 score of 1.925e-1. Both systems produced 0 false positives and 1,495 true negatives, resulting in a precision of 1 and an FPR of 0.

Overall, Slips performed better than Suricata at both evaluation levels because it achieved higher recall and F1 scores while maintaining the same negligible or zero false-positive rate. Its advantage was especially pronounced at the flow level, where it produced far more true positives than Suricata. At the profile-time-window level, Slips detected 33 malicious profile-time-window pairs, compared with 18 detected by Suricata. This indicates that Slips was more effective at identifying malicious behavior over time, which is consistent with the profile-time-window being its primary decision unit.

However, the recall of both systems remained low, meaning that neither provided comprehensive coverage of the malicious activity. The improvement observed for both systems at the profile-time-window level compared with their flow-by-flow results indicates that their detections are more effective for recognizing malicious host behavior accumulated across time than for labeling every individual malicious flow.

\begin{table*}[t]
\centering
\caption{Traffic detection comparison on Malware-1, Malware-2, and Benign-1. FPR is computed as \(FP/(FP+TN)\).}
\label{tab:malware-results}
\scriptsize
\begin{tabular}{llrrrrrrrr}
\toprule
Type & System & TP & FN & FP & TN & Precision & Recall & F1 & FPR \\
\midrule
Flow-by-flow & Slips & 3610 & 37834 & 2 & 78740 & 0.9994 & 0.0871 & \textbf{0.1603} & 0.000025 \\
Flow-by-flow & Suricata & 68 & 41376 & 2 & 78740 & 0.9714 & 0.0016 & 0.0033 & 0.000025 \\
Profile-window & Slips & 33 & 136 & 0 & 1495 & 1.0000 & 0.1953 & \textbf{0.3268} & 0.000000 \\
Profile-window & Suricata & 18 & 151 & 0 & 1495 & 1.0000 & 0.1065 & 0.1925 & 0.000000 \\
\bottomrule
\end{tabular}
\end{table*}

\section{Decision Case Study}
\label{sec:decision-case-study}

Alerts show whether Slips detected a malicious profile, but they do not show how the decision was made. We therefore trace one Slips alert from Malware-1, the Bladabindi RAT malware capture. The trace shows the complete path from a host-level alert to correlated evidence and then to the flow identifiers stored in the SQLite database.

The selected alert is reported for host \texttt{192.168.1.115} in \texttt{time window 2}. The run used Slips with a one-hour time window, analysis direction \texttt{out}, and an evidence threshold of 15 per hour. Slips created incident \texttt{749e232c-acee-4817-82f7-d71675ed1967} when the accumulated threat level for this profile and time window reached 15.0.

The incident contains a \texttt{CorrelID} list with 60 evidence identifiers. These correspond to 30 medium-severity evidence items and 30 informational evidence items. 

The medium evidence items report repeated connections from \texttt{192.168.1.115} to the unknown destination port \texttt{1177/TCP} on \texttt{41.108.179.197}. Each medium item has confidence 1.0 and contributes 0.5 to the accumulated score. 

The informational items indicate that the same destination IP was contacted without DNS resolution; they are retained for context but do not increase the score.

The evidence-to-flow link is explicit. For example, the evidence item \texttt{608dadb5-46a7-46f1-b348-4d9ea1379370}, which is one of the 60 evidence items reported by the above alert, points to flow UID \texttt{CJJC2G1hvW8XwV9cu3}.

Another is \texttt{7fb57255-3868-4ccf-96de-9ebbab62b548}, which points to flow UID \texttt{CzNzeW1yR9E17jomgg}. The latter flow is a TCP connection from \texttt{192.168.1.115:60308} to \texttt{41.108.179.197:1177}, with Zeek state \texttt{S0}, duration 9.009 s, three source packets, and no destination packets.

This chain lets an analyst move from the alert to the evidence and then to the concrete network observations.

This example uses heuristic evidence, but the same path is used by ML modules. An ML module emits evidence with a threat level, confidence, and flow identifiers; the evidence handler then combines it with other evidence for the same profile and time window. The final decision is therefore not an opaque model output. It is an accumulated host-level decision with stored links to the observations that supported it.

\input{previous-work-ids-architectures}

\section{Discussion and Limitations}
The architecture relies on shared state and calibrated evidence scores. IP-based profiles can merge devices behind address translation or split one device after an address change, while time-window boundaries can divide behavior. Correlated modules may detect overlapping aspects of the same behavior, causing related evidence to be counted multiple times and potentially inflating the resulting confidence and accumulated threat level score.
The illustrative evaluation does not assess the impact of these architectural limitations, Slips’ ability to analyze encrypted traffic, its resistance to adversarial evasion, its privacy implications, or its performance at operational scale.

\section{Conclusion}
Slips combines time-bounded host profiles, independent detection modules, and evidence-based decisions in one network security architecture. The design lets AI and non-AI modules contribute traceable findings without directly controlling the final alert. The evaluation illustrates this separation and its provenance, but supports conclusions only for the tested captures and configurations.

\bibliographystyle{ACM-Reference-Format}
\bibliography{references-slips,manual}

\section{Ethical Considerations}
\label{sec:ethics}

All network captures used in this paper were produced by the authors in a controlled laboratory environment under their exclusive control.
The benign capture records the authors' own browsing activity; the malicious captures were produced by executing malware samples and launching port scans on an isolated network segment.

\section{Generative AI Usage}
\label{sec:genai}

Generative AI tools were used to assist with software development/debugging and with manuscript structural flow and grammar. The authors are responsible for the research, validation, and final manuscript.

\appendix
\input{appendix-ids-comparison}

\input{appendix-open-science}

\end{document}

%% file: previous-work-ids-architectures.tex
\section{Previous Work: IDS Architectures}
\label{sec:previous-work-architectures}

We distinguish complete or prototype IDS architectures from classifiers evaluated only on benchmark rows without a defined acquisition, state, alert, or response pipeline. Table~\ref{tab:ids-architectures} compares systems that define an operational pipeline or explicit multi-component architecture. There, a flow means packets aggregated with defined key and time semantics, not merely one dataset row.

Production systems retain substantial multi-packet state. Snort~3 and Suricata combine decoding, stream or flow state, protocol inspection, rules, and actions~\cite{roesch_snort_1999,suricata_github}; Zeek converts packets into events that scripts correlate across connections~\cite{paxson_bro_1999}. They support stateful detections, but do not impose a detector-independent host profile and general evidence-accumulation decision.

Behavioral models also retain context, but usually for one model family. Kitsune maintains address- and channel-keyed statistics over several decay windows~\cite{mirsky_kitsune_2018}; DÏoT models ordered packet sequences per IoT device~\cite{nguyen_diot_2019}; and APChain compares chains of traffic attributes with a learned host model~\cite{seo_apchain_2018}. Their context is richer than one packet or flow, but their state and output remain specific to the corresponding detector.

Graph and correlation architectures combine lower-level observations. GrIDS reduces communication graphs, BotDet correlates Zeek-based detectors, and a peer-to-peer CIDS combines reports from several networks~\cite{cheung_grids,ghafir_botdet_2018,zhou_p2pcids_2005}. Their intermediate representations are task-specific rather than a general evidence object carrying threat, confidence, and supporting-flow references.

Model ensembles and modular pipelines provide another form of composition~\cite{hajizadeh_fsaids_2023,verkerken_hierarchical_2023,herrero_rtmovicab_2013,qiao_ainids_2002,alharbi_eds_2023}. They combine classifiers, agents, or IDS engines, but do not jointly provide shared host state, evidence accumulated over time, and provenance to supporting flows.

%% file: appendix-ids-comparison.tex
\section{Detailed IDS Architecture Comparison}
\label{app:ids-architectures}

Table~\ref{tab:ids-architectures} compares the architectures of selected prior intrusion detection systems.

\begin{table*}[t]
\centering
\caption{Architectural comparison of prior intrusion detection systems, ordered by creation year. ``Created'' refers to the system's first implementation or introduction rather than the publication year of the cited source.}
\label{tab:ids-architectures}
\footnotesize
\renewcommand{\arraystretch}{1.15}

\begin{tabularx}{\textwidth}{
    p{1.55cm}
    p{0.6cm}
    p{1.25cm}
    L
    p{1.05cm}
    L
    L
}
\toprule
System &
Created &
Input unit &
\makecell[l]{Multi-packet or\\multi-flow context} &
Profile &
\makecell[l]{Decision\\hierarchy} &
\makecell[l]{Ensemble/\\fusion} \\
\midrule

Zeek (Bro)~\cite{paxson_bro_1999} &
1995 &
Packets converted to typed protocol events &
Stateful connections, transactions, and scripts over event streams &
Script-defined host state &
Packet $\rightarrow$ protocol event $\rightarrow$ script notice/action &
Scripted correlation; no default weighted ensemble \\

GrIDS~\cite{cheung_grids} &
1996 &
Network activity reports &
Communication graphs, reductions, and administrative hierarchy &
Hosts as attributed graph entities &
Report $\rightarrow$ graph $\rightarrow$ rule match $\rightarrow$ alert &
Rule/report aggregation \\

Snort~\cite{roesch_snort_1999} &
1998 &
Packets and reconstructed streams &
Flow tracking, TCP reassembly, service inspection, and rule-local thresholds &
No shared behavioral host profile &
DAQ packet $\rightarrow$ codec/inspector $\rightarrow$ rule $\rightarrow$ action/output &
No general alert fusion \\

AINIDS~\cite{qiao_ainids_2002} &
2002 &
Packet headers &
Detector matches plus monitor-agent state &
Resource and detector state; not a host profile &
Match $\rightarrow$ co-stimulation $\rightarrow$ intrusion &
Heterogeneous danger signals \\

P2P CIDS~\cite{zhou_p2pcids_2005} &
2005 &
Suspicious-source reports &
Corroboration across subnetworks &
Minimal source-IP record &
Local report $\rightarrow$ distributed correlation $\rightarrow$ alert &
Multi-sensor agreement \\

Suricata~\cite{suricata_github} &
2007 &
Packets, flows, streams, and application transactions &
Bidirectional flow state, reassembly, application parsers, and rule-local thresholds &
No shared behavioral host profile &
Capture/decode $\rightarrow$ flow/stream/application layer $\rightarrow$ signature $\rightarrow$ alert/drop &
No general alert fusion \\

RT-MOVICAB-IDS~\cite{herrero_rtmovicab_2013} &
2013 &
TCP packet features &
Incremental learned cases; no explicit multi-flow host window &
No &
Projection $\rightarrow$ neural analysis $\rightarrow$ case-based decision/action &
Hybrid ANN and case-based reasoning \\

APChain~\cite{seo_apchain_2018} &
2018 &
Network-traffic attributes &
Chains of related host communications over time &
Per-host learned behavior model &
Traffic $\rightarrow$ chain $\rightarrow$ deviation $\rightarrow$ infected host &
No \\

BotDet~\cite{ghafir_botdet_2018} &
2018 &
Zeek connection and DNS events &
Per-host counters, DNS behavior, and event correlation &
Module-specific host tables &
Event $\rightarrow$ module detection $\rightarrow$ correlated bot alert &
Four heterogeneous detectors \\

Kitsune~\cite{mirsky_kitsune_2018} &
2018 &
Packets &
Damped address and channel statistics at multiple time scales &
Implicit address and address-pair statistics &
Packet $\rightarrow$ features $\rightarrow$ reconstruction errors $\rightarrow$ anomaly score &
Autoencoder ensemble \\

DÏoT~\cite{nguyen_diot_2019} &
2019 &
Packets encoded as symbols &
Ordered packet sequences &
Device type and learned communication model &
Packet $\rightarrow$ sequence score $\rightarrow$ device alert &
Federated model aggregation \\

FSA-IDS~\cite{hajizadeh_fsaids_2023} &
2023 &
Argus five-tuple flows &
Packets only within each flow &
No &
Flow $\rightarrow$ ensemble uncertainty $\rightarrow$ pseudo-label/expert query &
Supervised classifier ensemble \\

Hierarchical IDS~\cite{verkerken_hierarchical_2023} &
2023 &
Benchmark flow records &
No context across flows &
No &
Benign/suspicious $\rightarrow$ class/unknown &
No; serial classifiers \\

EDS~\cite{alharbi_eds_2023} &
2023 &
Traffic plus outputs from three IDS engines &
Delegated to constituent IDSs &
Searchable events; no unified profile &
Detector output $\rightarrow$ SIEM index/dashboard &
Parallel IDS aggregation \\

\bottomrule
\end{tabularx}
\end{table*}

%% file: appendix-open-science.tex
\section{Open Science Appendix}
\label{app:open-science}

All datasets, tools, experimental results, and supporting artifacts used in this paper have been anonymized and made publicly available to support reproducibility. Because the complete artifact package exceeds GitHub's recommended repository and file size limits, it has been compressed and hosted in an anonymized Zenodo repository:

\begin{center}
\url{https://zenodo.org/records/21344815}
\end{center}